\documentclass[12pt,preprint]{aastex}

\slugcomment{to appear in the \emph{Astrophysical Journal}}

\shorttitle{Water and Oxygen Abundances on Saturn}

\shortauthors{Visscher and Fegley}

\begin{document}

\title{Chemical Constraints on the Water and Total Oxygen Abundances in the Deep Atmosphere of Saturn}

\author{Channon Visscher and Bruce Fegley, Jr.}
\affil{Planetary Chemistry Laboratory, Department of Earth \&
Planetary Sciences, McDonnell Center for the Space Sciences,
Washington University, St. Louis, MO 63130-4899}

\email{visscher@wustl.edu, bfegley@wustl.edu}

\begin{abstract}
Thermochemical equilibrium and kinetic calculations for the trace
gases CO, PH$_{3}$, and SiH$_{4}$ give three independent
constraints on the water and total oxygen abundances of Saturn's
deep atmosphere. A lower limit to the water abundance of
H$_{2}$O/H$_{2}$ $\geq$ (1.7$_{-0.4}^{+0.7}$)$\times$10$^{-3}$ is
given by CO chemistry while an upper limit of H$_{2}$O/H$_{2}$
$\leq$ (5.5$_{-2.5}^{+0.8}$)$\times$10$^{-3}$ is given by PH$_{3}$
chemistry.  A combination of the CO and PH$_{3}$ constraints
indicates a water enrichment on Saturn of 1.9 to 6.1 times the
solar system abundance (H$_{2}$O/H$_{2}$ = 8.96$\times$10$^{-4}$).
The total oxygen abundance must be at least 1.7 times the solar
system abundance (O/H$_{2}$ = 1.16$\times$10$^{-3}$) in order for
SiH$_{4}$ to remain below the detection limit of SiH$_{4}$/H$_{2}$
\textless 2$\times10^{-10}$.  A combination of the CO, PH$_{3}$,
and SiH$_{4}$ constraints suggests that the total oxygen abundance
on Saturn is 3.2 to 6.4 times the solar system abundance.  Our
results indicate that oxygen on Saturn is less enriched than other
heavy elements (such as C and P) relative to the solar system
composition.
\end{abstract}


\pagebreak

\section{Introduction}

Spectroscopic observations of water in the upper atmosphere of
Saturn indicate a H$_{2}$O/H$_{2}$ mixing ratio (\emph{q}H$_{2}$O)
of $\sim$(2-200)$\times$10$^{-9}$ (Larson et al. 1980; Winkelstein
et al. 1983; Chen et al. 1991; de Graauw et al. 1997; Feuchtgruber
et al. 1997), well below the solar system (i.e., protosolar)
H$_{2}$O/H$_{2}$ ratio of $\sim9\times10^{-4}$. Measurements of
H$_{2}$O in the 5 $\mu$m window probe the $\sim3$ bar level in
Saturn's troposphere (Larson et al. 1980; de Graauw et al. 1997).
However, H$_{2}$O cloud condensation occurs deeper, near the 13
bar level, so the water abundance below the clouds and therefore
the planetary inventory of water remains unknown. Observations of
Saturn's atmosphere show that CH$_{4}$/H$_{2}$, PH$_{3}$/H$_{2}$,
and AsH$_{3}$/H$_{2}$ ratios are enhanced over protosolar values,
suggesting a similar enrichment may exist for water.  The total
oxygen abundance of Saturn's interior is expected to primarily
consist of H$_{2}$O and O bound in rock. Since oxygen is the third
most abundant element in the solar system, water vapor is expected
to be an important gas below the clouds of Saturn.

Here we consider the effects of water and oxygen on tropospheric
chemistry to determine the water and total oxygen abundances of
Saturn's deep atmosphere.  We specifically focus on the
disequilibrium trace gases CO, PH$_{3}$, and SiH$_{4}$ because
their chemistry is sensitive to the water and oxygen content of
the troposphere. Using a similar approach as Fegley \& Prinn
(1988), we show that the observed abundance of CO and the upper
limit for SiH$_{4}$ are incompatible with significant planetary
depletions in water and oxygen, while the PH$_{3}$ abundance is
incompatible with large enrichments. Taken together, our results
indicate that water and total oxygen on Saturn are less enriched
than heavy elements such as C and P relative to solar system
composition. We begin with an overview of the observed composition
of Saturn's atmosphere (\S\ref{Atmospheric Composition of Saturn})
and a brief description (\S\ref{Method}) of our computational
method. In \S\ref{Results}, we present an overview of Saturn's
atmospheric chemistry and our results for the CO, PH$_{3}$, and
SiH$_{4}$ chemical constraints, followed by discussion
(\S\ref{Discussion}) of their implications for the water and
oxygen abundances of Saturn's interior.  A summary is given in
\S\ref{Summary}.

\section{Atmospheric Composition of Saturn}\label{Atmospheric Composition of Saturn}

The observed mixing ratios for several compounds in Saturn's
atmosphere are listed in Table 1, along with computed enrichments
over solar system abundances.  Methane, PH$_{3}$, and AsH$_{3}$
are the major C-, P-, As-bearing gases in Saturn's atmosphere
(Lodders \& Fegley 1998). It is generally assumed that the
CH$_{4}$, PH$_{3}$, and AsH$_{3}$ abundances represent the total
elemental abundances of C, P, and As, respectively, in Saturn's
observable atmosphere (e.g., Courtin et al. 1984, Noll et al.
1989, Noll \& Larson 1990, B\'ezard et al. 1989, Fegley \& Lodders
1994; Hersant et al. 2004), and the same approach is taken here.
Saturn's atmospheric composition below the clouds is generally
considered to be uniform because of convective mixing. However,
the distribution of elements between different gases (e.g.,
H$_{2}$O, CO, OH) and (in some cases) between gases and
condensates (e.g., H$_{2}$O, rock) is temperature and pressure
dependent.  Although similar within observational uncertainties,
the enrichment factor for As is apparently less than that for C
and P, possibly due to the formation of other As-bearing gases
such as AsF$_{3}$ (see Fegley \& Lodders 1994). Ammonia is almost
certainly the major N-bearing gas in Saturn's atmosphere. However,
the NH$_{3}$ abundance is affected by cloud condensation and
photolysis and cannot be used as a constraint on the nitrogen
enrichment relative to protosolar composition. Likewise, the
observed abundances of other gases such as H$_{2}$S and GeH$_{4}$
cannot be used as constraints because they are affected by
condensation (solid NH$_{4}$SH, Ge, GeTe), photolysis (H$_{2}$S),
and formation of other gases (GeS, GeSe, GeTe) (Fegley \& Lodders
1994).  Briggs \& Sackett (1989) inferred a H$_{2}$S abundance on
Saturn of $\sim 10$ times the protosolar value. However, because
it is difficult to distinguish the microwave opacity of H$_{2}$S
from other sources, this H$_{2}$S abundance is an indirect
estimate based on the brightness temperature spectrum for an
assumed NH$_{4}$SH cloud (Hersant et al. 2004). Thus at present we
only use the observed enrichments in C and P to constrain the
average enrichment of heavy elements on Saturn relative to the
solar system composition.

Elemental abundances for the solar nebula (i.e., protosolar
abundances) were taken from Lodders (2003).  These are slightly
different from photospheric abundances due to heavy element
settling in the Sun (Lodders 2003).  The protosolar elemental
abundances represent the \emph{bulk} elemental composition of the
Sun and the solar nebula.  Water vapor is expected to be the
dominant O-bearing gas in the circum-Saturnian nebula and in
Saturn's deep atmosphere (e.g., see \S\ref{Overview}; Fegley \&
Prinn 1985; 1989), and the CO/H$_{2}$O ratio in both environments
is much less than unity. The protosolar H$_{2}$O/H$_{2}$ ratio is
defined by taking the total oxygen abundance
(O/H$_{2}=1.16\times10^{-3}$) and subtracting the portion that
forms rock (see Lodders 2004). This is expressed as
\begin{equation}\label{oxygenmassbalance}
\textrm{O}_{\textrm{\scriptsize{H}}_{2}\textrm{\scriptsize{O}}}=\Sigma\textrm{O}-\textrm{O}_{\textrm{\scriptsize{rock}}}
\end{equation}
where the amount of oxygen bound in rock
(MgO+SiO$_{2}$+CaO+Al$_{2}$O$_{3}$+Na$_{2}$O+K$_{2}$O+TiO$_{2}$) is
given by
\begin{equation}\label{Orock}
\textrm{O}_{\textrm{\scriptsize{rock}}}=(\textrm{Mg+2Si+Ca+1.5Al+0.5Na+0.5K+2Ti}).
\end{equation}
In a gas with protosolar elemental abundances, the formation of
rock effectively removes $\sim$23\% of the total oxygen.
Throughout the following, enrichments in water over the protosolar
composition refer to a solar system H$_{2}$O/H$_{2}$ ratio of
$8.96\times10^{-4}$.

\section{Method}\label{Method}

Thermochemical equilibrium calculations were performed using a
Gibbs free energy minimization code and an adiabatic
temperature-pressure profile for Saturn's troposphere calculated
as described by Fegley \& Prinn (1985) using \emph{T} = 134.8 K at
\emph{P} = 1 bar (Lindal et al. 1985) and a total He/H$_{2}$ ratio
of 0.135.  This value is the mean of the volume mixing ratio
He/H$_{2}$ = 0.11-0.16 determined by Conrath \& Gautier (2000),
considerably greater than the previously accepted value of
0.034$\pm$0.024 (Conrath et al. 1984). The corresponding mole
fraction of hydrogen (\emph{X}$_{\textrm{\scriptsize{H}}_{2}}$) is
0.881. We adopted a nominal enrichment factor of 7.4 times the
protosolar element/H$_{2}$ ratios for elements heavier than He
based upon the observed enrichments of CH$_{4}$ and PH$_{3}$ on
Saturn, as described in \S\ref{Atmospheric Composition of Saturn}
and shown in Table 1. We also varied the elemental abundances of
C, P, Si, and O in order to study the resulting effects on
Saturn's tropospheric chemistry.

\section{Results}\label{Results}

\subsection{Overview of Atmospheric Chemistry}\label{Overview}

The model adiabatic profile for Saturn's atmosphere is shown in
Figure 1. This figure also shows the results of thermochemical
equilibrium calculations for a gas with protosolar elemental
abundances. The lines labelled Fe (s,l), Mg$_{2}$SiO$_{4}$ (s,l),
and MgSiO$_{3}$ (s,l) are the condensation curves for iron,
forsterite (Mg$_{2}$SiO$_{4}$), and enstatite (MgSiO$_{3}$), with
open circles denoting their normal melting points. These three
phases constitute most of the ``rock'' that is expected to
condense in Saturn's deep atmosphere.  The curves labelled
CO/CH$_{4}$, N$_{2}$/NH$_{3}$, and PH$_{3}$/P$_{4}$O$_{6}$ show
where the partial pressures of these gases are equal, and they are
interpreted as follows. Methane (CH$_{4}$) is the major
carbon-bearing gas to the right of the CO/CH$_{4}$ curve, and
carbon monoxide (CO) is the major carbon-bearing gas to the left
of the CO/CH$_{4}$ curve. However, CH$_{4}$ is still present, but
is less abundant than CO, inside the CO field and vice versa.
Likewise ammonia (NH$_{3}$) is the major nitrogen-bearing gas to
the right of the N$_{2}$/NH$_{3}$ curve, and molecular nitrogen
(N$_{2}$) is the major nitrogen-bearing gas to the left of the
N$_{2}$/NH$_{3}$ curve. Ammonia is still present, but is less
abundant than N$_{2}$, inside the N$_{2}$ field, and vice versa. A
comparison of the model Saturnian adiabat with the CO/CH$_{4}$ and
N$_{2}$/NH$_{3}$ curves shows that methane and ammonia are
predicted to be the dominant C-bearing and N-bearing gases
throughout Saturn's atmosphere at temperatures below 3000 K.

At pressures characteristic of the deep atmospheres of Saturn and
the other gas giant planets, phosphine (PH$_{3}$) is the major
phosphorus-bearing gas to the top of the PH$_{3}$/P$_{4}$O$_{6}$
curve, and P$_{4}$O$_{6}$ gas is the major phosphorus-bearing gas
to the bottom of the PH$_{3}$/P$_{4}$O$_{6}$ curve. The partial
pressures of the two gases are equal at about the 900 K level in
Saturn's atmosphere. This curve is extrapolated to lower pressures
but at these pressures phosphorus chemistry becomes more complex
and a number of phosphorus-bearing molecules are found in the gas
at high temperatures (e.g., Fegley and Lewis 1980).

\subsection{Carbon Monoxide}\label{Carbon Monoxide}

We first consider carbon monoxide (CO), which is observed in
Saturn's atmosphere at a mixing ratio (CO/H$_{2}$) of
$(1.6\pm0.8)\times10^{-9}$ (Noll et al. 1986; Noll \& Larson 1990;
de Graauw et al. 1997).  This is $\sim$40 orders of magnitude
higher than the CO abundance predicted by thermodynamic
equilibrium in Saturn's cool, visible atmosphere (e.g., see Fegley
\& Prinn 1985; Fegley \& Lodders 1994). As discussed by Fegley \&
Prinn (1985, p. 1076), CO in Saturn's observable atmosphere may
result from a combination of internal and external sources.  We
consider the effects of an additional external source in \S
\ref{External CO Source}. Carbon monoxide in Saturn's deep
atmosphere is produced from water via the net thermochemical
reaction
\begin{equation}\label{COeq}
\textrm{CH}_{4}+\textrm{H}_{2}\textrm{O}=\textrm{CO}+3\textrm{H}_{2}
\end{equation}
The corresponding equilibrium constant expression for reaction
(\ref{COeq}) is
\begin{equation}\label{COKeq}
K_{\ref{COeq}}=[(X_{\textrm{\scriptsize{CO}}}X_{\textrm{\scriptsize{H}}_{2}}^{3})/(X_{\textrm{\scriptsize{CH}}_{4}}X_{\textrm{\scriptsize{H}}_{2}\textrm{\scriptsize{O}}})]P_{T}^{2}
\end{equation}
where $K_{\ref{COeq}}$ is the equilibrium constant and $P_{T}$ is
the total pressure along the adiabatic profile.  Rearranging
equation (\ref{COKeq}) and substituting mixing ratios for mole
fractions for CO, CH$_{4}$, and H$_{2}$O (e.g.,
$q\textrm{CO}=X_{\textrm{\scriptsize{CO}}}/X_{\textrm{\scriptsize{H}}_{2}}$),
the CO mixing ratio is given by
\begin{equation}\label{qCO}
q\textrm{CO}=(q\textrm{CH}_{4}q\textrm{H}_{2}\textrm{O}/X_{\textrm{\scriptsize{H}}_{2}}^{2})K_{\ref{COeq}}P_{T}^{-2}
\end{equation}
We now rewrite equation (\ref{qCO}) to explicitly show the
dependence of the CO mixing ratio upon the CH$_{4}$ and H$_{2}$O
enrichments relative to solar system composition.  The protosolar
mixing ratios for methane ($5.82\times10^{-4}$) and water
($8.96\times10^{-4}$) and the hydrogen mole fraction of
\emph{X}$_{\textrm{\scriptsize{H}}_{2}}=0.881$ are constants,
allowing us to write
\begin{equation}\label{c'}
c'=(q\textrm{CH}_{4}q\textrm{H}_{2}\textrm{O})_{\textrm{\scriptsize{protosolar}}}/X_{\textrm{\scriptsize{H}}_{2}}^{2}=6.72\times10^{-7}
\end{equation}
Substitution into equation (\ref{qCO}) gives
\begin{equation}\label{qCOsolar}
q\textrm{CO}=c'E_{\textrm{\scriptsize{CH}}_{4}}E_{\textrm{\scriptsize{H}}_{2}\textrm{\scriptsize{O}}}K_{\ref{COeq}}P_{T}^{-2}
\end{equation}
where $E_{\textrm{\scriptsize{CH}}_{4}}$ and
$E_{\textrm{\scriptsize{H}}_{2}\textrm{\scriptsize{O}}}$ are
enrichment factors over the protosolar composition for the methane
(i.e., carbon) and water abundances in Saturn's atmosphere.
Examination of equation (\ref{qCOsolar}) shows that the CO mixing
ratio is proportional to the product
($E_{\textrm{\scriptsize{CH}}_{4}}$$E_{\textrm{\scriptsize{H}}_{2}\textrm{\scriptsize{O}}}$).
 Thus for otherwise constant conditions, the water enrichment
required to produce a given CO abundance varies inversely with the
methane enrichment.  The equilibrium CO mixing ratio as a function
of temperature (expressed as $\xi=10^{4}/(T,\textrm{K})$) along
the Saturnian adiabat is given by the equation
\begin{eqnarray}\label{COfit}
\log q\textrm{CO}=\log
c'-29.7374-1.1770\xi^{-1}-1.509\times10^{-3}\xi^{2}-34.3300\xi^{1/2}
\nonumber\\+0.1619\xi-1.898\times10^{-5}\xi^{3}+61.5475\xi^{1/3}+\log
E_{\textrm{\scriptsize{CH}}_{4}}+\log
E_{\textrm{\scriptsize{H}}_{2}\textrm{\scriptsize{O}}}
\end{eqnarray}
from 300-6000 K, where $\log c'=-6.1726$ from equation (\ref{c'}).
Equation (\ref{COfit}) has the form of a heat capacity polynomial
to account for the strong temperature dependence of the
equilibrium constant $K_{\ref{COeq}}$.

\subsubsection{Thermodynamic Limit}

At constant pressure, reaction (\ref{COeq}) proceeds to the right
with increasing temperature and yields more CO (Lodders \& Fegley
2002).  However, at constant temperature, reaction (\ref{COeq})
proceeds to the left with increasing pressure and yields less CO. In
other words, the equilibrium abundance of CO increases with
increasing temperature and with decreasing pressure.  Therefore a
maximum occurs in the CO mixing ratio along the Saturnian adiabat,
found by differentiating equation (\ref{COfit}) and solving for the
temperature at which the derivative is zero:
\begin{eqnarray}
(d \log
q\textrm{CO}/d\xi)=0=1.1770\xi^{-2}-3.018\times10^{-3}\xi-17.1650\xi^{-1/2}
\nonumber\\+ 0.1619-5.694\times10^{-5}\xi^{2}+20.5158\xi^{-2/3}
\end{eqnarray}
which gives the maximum at 2910 K.  The greatest CO abundance is
achieved at this temperature over a wide range of water
enrichments and therefore it serves as a thermodynamic lower limit
to the total water abundance in Saturn's interior.  Using $T=2910$
K and the observed mixing ratio
$q\textrm{CO}=(1.6\pm0.8)\times10^{-9}$ (see Table 1) in equation
(\ref{COfit}) gives
\begin{equation}
\log E_{\textrm{\scriptsize{CH}}_{4}}+\log
E_{\textrm{\scriptsize{H}}_{2}\textrm{\scriptsize{O}}}=\log
q\textrm{CO}+6.476=-2.320_{-0.301}^{+0.176}
\end{equation}
At the nominal carbon enrichment of of 7.4 times protosolar, the
enrichment factor for water is
$E_{\textrm{\scriptsize{H}}_{2}\textrm{\scriptsize{O}}}=(6.5_{-3.3}^{+3.2}
)\times10^{-4}$. Using the lower bound of this value indicates
that
$E_{\textrm{\scriptsize{H}}_{2}\textrm{\scriptsize{O}}}\geq3.2\times10^{-4}$
is necessary to produce the observed CO abundance, corresponding
to a H$_{2}$O/H$_{2}$ mixing ratio of $2.9\times10^{-7}$ in
Saturn's troposphere. Thus, water cannot be depleted more than
$3.2\times10^{-4}$ times the protosolar value if the observed
amount of CO is produced in Saturn's interior. This thermodynamic
constraint gives a firm lower limit to the amount of water in
Saturn's deep atmosphere.  However, it implies mixing of gas from
an unrealistic depth (the 2910 K, 46 kbar level) up to the visible
atmosphere. In order to better constrain a lower limit to the
water abundance, the kinetics of CO destruction must be
considered.

\subsubsection{Kinetic Limit}\label{Kinetic Limit}

As parcels of hot gas rise in Saturn's atmosphere, CO is destroyed
by conversion to CH$_{4}$.  Therefore the observable amount of CO
depends on both the rate of vertical mixing and the kinetics of
conversion (e.g., Fegley \& Prinn 1985). In the kinetic scheme
proposed by Prinn \& Barshay (1977), CO is in equilibrium with
formaldehyde and the rate-limiting step for CO destruction is the
breaking of the C$\dbond$O bond in formaldehyde via
\begin{equation}\label{PB}
\textrm{H}_{2}\textrm{CO}+\textrm{H}_{2}\longrightarrow\textrm{CH}_{3}+\textrm{OH}
\end{equation}
Assuming that the maximum plausible rate of vertical mixing is
given by $K_{eddy}\sim10^{9}$ cm$^{2}$ s$^{-1}$, estimated from
Saturn's internal heat flux (see Prinn et al. 1984, p. 138), CO
destruction is quenched at the 1036 K level on Saturn.  Using
$T=1036$ K in equation (\ref{COfit}) along with the observed CO
abundance gives
\begin{equation}\label{kineticwaterlimit}
\log E_{\textrm{\scriptsize{CH}}_{4}}+\log
E_{\textrm{\scriptsize{H}}_{2}\textrm{\scriptsize{O}}}=\log
q\textrm{CO}+10.239=1.443_{-0.301}^{+0.176}
\end{equation}
Equation (\ref{kineticwaterlimit}) defines the lower limit for
enrichments in the carbon and water abundances on Saturn.  At the
nominal carbon enrichment of 7.4 times protosolar, the enrichment
factor for water is
$E_{\textrm{\scriptsize{H}}_{2}\textrm{\scriptsize{O}}}=3.7_{-1.8}^{+1.9}$.
The lower bound of this value indicates that
$E_{\textrm{\scriptsize{H}}_{2}\textrm{\scriptsize{O}}}\geq1.9$ is
required to produce the observed CO abundance.  This corresponds
to a H$_{2}$O/H$_{2}$ mixing ratio of \emph{q}H$_{2}$O
$\geq1.7\times10^{-3}$ in Saturn's deep atmosphere.

\subsubsection{Alternative Thermochemical Kinetics}

An alternative kinetic scheme for CO destruction was proposed by
Yung et al. (1988) where the rate limiting step involves the
conversion of the C$\dbond$O bond in formaldehyde into a C$\sbond$O
bond via
\begin{equation}\label{Yung}
\textrm{H}+\textrm{H}_{2}\textrm{CO}+\textrm{M}\longrightarrow\textrm{CH}_{3}\textrm{O}+\textrm{M}
\end{equation}
This alternative kinetic scheme gives significantly less CO at the
same vertical mixing rate than the Prinn \& Barshay (1977) model
because the rate determining step is significantly faster (e.g.,
see Yung et al. 1988 and B\'ezard et al. 2002).  Vertical mixing
that is orders of magnitude more rapid than implied by observed
heat fluxes or mixing lengths significantly smaller than pressure
scale heights are required to match observed CO abundances on
Jupiter and Saturn using this alternative kinetic scheme.  Leaving
these problems aside, we compute a quench temperature for reaction
(\ref{Yung}) of 816 K on Saturn.  In this case, using $T=816$ K in
equation (\ref{COfit}) with the observed CO mixing ratio gives
\begin{equation}\label{Bezardkineticwaterlimit}
\log E_{\textrm{\scriptsize{CH}}_{4}}+\log
E_{\textrm{\scriptsize{H}}_{2}\textrm{\scriptsize{O}}}=\log
q\textrm{CO}+12.563=3.767_{-0.301}^{+0.176}
\end{equation}
The nominal $E_{\textrm{\scriptsize{CH}}_{4}}$ value of 7.4 times
protosolar requires
$E_{\textrm{\scriptsize{H}}_{2}\textrm{\scriptsize{O}}}\geq395$,
or a water abundance of $q\textrm{H}_{2}\textrm{O}\geq0.35$ in
Saturn's atmosphere. If we employ a mixing length of $L\sim0.1H$
in place of the pressure scale height $H$ (B\'ezard et al. 2002,
Smith 1998), the quench temperature is 922 K, which requires
$E_{\textrm{\scriptsize{H}}_{2}\textrm{\scriptsize{O}}}\geq20.9$,
and a water abundance of
$q\textrm{H}_{2}\textrm{O}\geq1.9\times10^{-2}$ in Saturn's
atmosphere. However, as we discuss in \S \ref{Phosphine}, the
observed PH$_{3}$ abundance precludes a water enrichment that is
this large. A detailed mechanistic examination of CO quenching
kinetics is beyond the scope of this paper.  Here we continue our
use of the Prinn \& Barshay (1977) kinetic scheme for CO
destruction kinetics because it accurately reproduces the observed
CO abundance on Jupiter, Saturn, and Neptune (Fegley \& Lodders
1994; Lodders \& Fegley 1994).

\subsubsection{External CO Source}\label{External CO Source}

As suggested by Fegley \& Prinn (1985), Saturn may also have a
competing external source of CO.  If present, an external CO source
would lower the amount of tropospheric water required to produce the
observed CO abundance.  Possible external sources include direct
delivery of CO or photolytic production via stratospheric water from
interplanetary dust, cometary impacts, or infalling ring or
satellite debris (e.g., Fegley \& Prinn 1985; Moses et al. 2000).

Measurements of carbon monoxide on Saturn suggest, but are not
diagnostic of, a primarily internal source that produces \emph{q}CO
$\sim10^{-9}$ in the upper troposphere (e.g., see Noll et al. 1986;
Noll \& Larson 1990; Moses et al. 2000; Ollivier et al. 2000), and
most spectroscopic models for the observed CO abundance on Saturn
include both an internal and external source. However, the relative
strength of each source is currently unknown. We therefore
considered scenarios which include both an internal and external
source of CO.

Results are shown in Figure 2 for internal CO fluxes comprising
100\%, 50\%, and 10\% of the lower bound of the observed CO mixing
ratio. At higher quench temperatures ($T_{quench}$), CO originates
deeper in the troposphere where it is thermodynamically more
stable and thus smaller water enrichments are required to produce
its observed abundance.  At lower quench temperatures, CO
originates higher in the troposphere where it is thermodynamically
less stable and therefore requires larger water enrichments to
achieve the observed abundance. However, the upper limit on the
water enrichment given by PH$_{3}$ chemistry (see
\S\ref{Phosphine}) indicates that some of the observed CO in
Saturn's atmosphere must come from an external source if the CO
quench temperature is less than 977 K.

At a constant quench temperature, less tropospheric water is
required if some of the observed CO is external in origin.  For
example, in their model distribution \textbf{S} in Noll \& Larson
(1990) the observed CO is mostly stratospheric (external), and
$q\textrm{CO}=10^{-10}$ in the troposphere, or about 10\% of the
nominal CO abundance.  At a quench temperature of 1036 K, model
\textbf{S} requires a tropospheric water enrichment of
$E_{\textrm{\scriptsize{H}}_{2}\textrm{\scriptsize{O}}}\geq0.2$,
compared to
$E_{\textrm{\scriptsize{H}}_{2}\textrm{\scriptsize{O}}}\geq1.9$
when all of the observed CO comes from an internal source.  For
our present discussion, we assume a primarily internal source that
produces a mixing ratio of $q\textrm{CO}=(1.6\pm0.8)\times10^{-9}$
and thus requires a water abundance of
H$_{2}$O/H$_{2}\geq1.7\times10^{-3}$ in Saturn's troposphere (see
\S\ref{Kinetic Limit}).

\subsection{Phosphine}\label{Phosphine}

Phosphine (PH$_{3}$) is observed in Saturn's atmosphere at a
mixing ratio (PH$_{3}$/H$_{2}$) of (5.1$\pm$1.6)$\times$10$^{-6}$
(Table 1). Because it is destroyed by tropospheric water, PH$_{3}$
constrains the upper limit of Saturn's water abundance. The
phosphine abundance is governed by the net thermochemical
equilibrium
\begin{equation}\label{PH3eq}
4\textrm{PH}_{3}+6\textrm{H}_{2}\textrm{O}=\textrm{P}_{4}\textrm{O}_{6}+12\textrm{H}_{2}
\end{equation}
which shows that, according to LeCh\^{a}telier's principle, the
PH$_{3}$ abundance decreases as the water abundance increases to
maintain chemical equilibrium.  Rearranging the equilibrium constant
expression for reaction (\ref{PH3eq}), the phosphine abundance is
given by
\begin{equation}\label{XPH3}
X_{\textrm{\scriptsize{PH}}_{3}}=[(X_{\textrm{\scriptsize{P}}_{4}\textrm{\scriptsize{O}}_{6}}X_{\textrm{\scriptsize{H}}_{2}}^{12}P_{T}^{3})/(X_{\textrm{\scriptsize{H}}_{2}\textrm{\scriptsize{O}}}^{6}K_{\ref{PH3eq}})]^{1/4}
\end{equation}
At high temperatures and/or low water abundances, phosphine is the
dominant P-bearing gas and \emph{q}PH$_{3}\approx q\Sigma$P.  As
parcels of hot gas rise in Saturn's atmosphere, PH$_{3}$ is
oxidized to P$_{4}$O$_{6}$.  The observed abundance of PH$_{3}$ is
$\sim30$ orders of magnitude higher than that predicted by
thermodynamic equilibrium and its presence is evidence of rapid
vertical mixing from Saturn's deep atmosphere (Fegley \& Prinn
1985).  Unlike CO, phosphine gives no inherent thermodynamic limit
to water enrichment because the PH$_{3}$ abundance generally
increases with both temperature and pressure (cf. Figure 3, Fegley
\& Prinn 1985). Therefore we turn directly to PH$_{3}$ quenching
kinetics to constrain the upper limit on the water abundance.

The observable amount of phosphine depends on both the rate of
vertical mixing and the kinetics of PH$_{3}$ destruction.  Prinn
et al. (1984) proposed a mechanism for PH$_{3}$ destruction where
the rate-determining step is formation of the P$\sbond$O bond via:
\begin{equation}\label{PrinnPH3}
\textrm{PH}+\textrm{OH}\longrightarrow\textrm{PO}+\textrm{H}_{2}
\end{equation}
Based on work by Twarowski (1995) we take the formation of the
P$\sbond$O bond by direct reaction of PH$_{3}$ with an OH radical as
our rate-determining step:
\begin{equation}\label{TwarowskiPH3}
\textrm{PH}_{3}+\textrm{OH}\longrightarrow\textrm{H}_{2}\textrm{POH}+\textrm{H}
\end{equation}
The corresponding chemical lifetime for PH$_{3}$ is
\begin{equation}\label{PH3lifetime}
t_{chem}(\textrm{PH}_{3})=1/(k_{\ref{TwarowskiPH3}}[\textrm{OH}])
\end{equation}
where the estimated rate constant $k_{\ref{TwarowskiPH3}}$ is
obtained from Twarowski's (1995) kinetic study of phosphine
combustion products and is given by
\begin{equation}
k_{\ref{TwarowskiPH3}}\approx5.25\times10^{-13}\exp(-6013.6/T)\textrm{
cm}^{3}\textrm{ s}^{-1}
\end{equation}
We again assume that the maximum plausible rate of vertical mixing
is given by $K_{eddy}\sim10^{9}$ cm$^{2}$ s$^{-1}$, and find that
using reaction ($\ref{TwarowskiPH3}$) as the rate-determining step
for phosphine destruction gives similar results as reaction
($\ref{PrinnPH3}$). Solving equations ($\ref{XPH3}$) and
($\ref{PH3lifetime}$) at a phosphorus enrichment of $7.4\pm2.3$
times protosolar shows that H$_{2}$O in Saturn's deep atmosphere
cannot be enriched more than $6.1_{-2.7}^{+0.9}$ times the solar
system abundance because greater water enrichments would reduce
the PH$_{3}$ abundance below the observed level. This water
enrichment corresponds to a H$_{2}$O/H$_{2}$ mixing ratio of
$q\textrm{H}_{2}\textrm{O}\leq(5.5_{-2.5}^{+0.8})\times10^{-3}$ in
Saturn's troposphere.

\subsection{Silane}

Silane (SiH$_{4}$) is destroyed by water in the deep atmosphere of
Saturn and serves as a constraint on the total oxygen abundance of
Saturn's interior.  Silicon is about 120, 8,300, and 164,000 times
more abundant than P, Ge, and As, respectively, in protosolar
composition gas. However, while PH$_{3}$, GeH$_{4}$, and AsH$_{3}$
have each been observed on Saturn (see Table 1), SiH$_{4}$
(predicted to be the dominant Si-bearing gas) remains undetected
with an upper limit of \emph{q}{SiH}$_{4}$\textless
$(0.2-1.2)\times10^{-9}$ (Larson et al. 1980; Noll \& Larson
1990). This is because silicon is efficiently removed from
Saturn's atmosphere by the formation of silicates such as
MgSiO$_{3}$ (enstatite) and Mg$_{2}$SiO$_{4}$ (forsterite),
exemplified by the net thermochemical reaction (Fegley \& Prinn
1988):
\begin{equation}\label{silaneoxidation}
\textrm{SiH}_{4}+2\textrm{H}_{2}\textrm{O}=4\textrm{H}_{2}+\textrm{SiO}_{2}(\textrm{solid,
liquid})
\end{equation}
which incorporates Si into rock.  In order for reaction
(\ref{silaneoxidation}) to remove Si from the atmosphere it is
evident that water vapor must be present to oxidize SiH$_{4}$.  This
is only possible if the total oxygen abundance ($\Sigma$O) is
greater than or equal to the oxygen consumed by rock-forming oxides
(O$_{\textrm{\scriptsize{rock}}}$). This mass balance constraint can
be written as
\begin{equation}\label{Simassbalanceconstraint}
\Sigma\textrm{O}/\textrm{O}_{\textrm{\scriptsize{rock}}}\geq1
\end{equation}
where the amount of oxygen bound in rock is given by equation
(\ref{Orock}).  Protosolar composition gas (Lodders 2003) has
\begin{equation}\label{Simassbalancesolar}
\Sigma\textrm{O}/\textrm{O}_{\textrm{\scriptsize{rock}}}=4.4
\end{equation}
which clearly satisfies the mass balance criterion.  Thus in order
for reaction (\ref{silaneoxidation}) to destroy all SiH$_{4}$ on
Saturn, the oxygen enrichment ($E_{\textrm{\scriptsize{O}}}$) in
Saturn's deep atmosphere must be
\begin{equation}\label{Simassconstraintsolar}
E_{\textrm{\scriptsize{O}}} \geq E_{\textrm{\scriptsize{rock}}}/4.4
\end{equation}
where
$E_{\textrm{\scriptsize{rock}}}=E_{\textrm{\scriptsize{Mg}}}=
E_{\textrm{\scriptsize{Si}}}$, etc., that is, all rock-forming
elements are assumed to be equally enriched. This assumption is
not necessarily correct because it appears that P and As, both of
which are rock-forming elements, may not be equally enriched on
Saturn (e.g., Noll et al. 1989; Noll \& Larson 1990; see Table 1
and \S\ref{Atmospheric Composition of Saturn}). However, we
explicitly assume equal enrichment of rock-forming elements and
take $E_{\textrm{\scriptsize{rock}}}=7.4\pm2.3$ by comparison with
the observed enrichment in phosphorus. Equation
(\ref{Simassconstraintsolar}) then gives a total oxygen enrichment
of $E_{\textrm{\scriptsize{O}}}\geq1.7\pm0.5$ times the protosolar
oxygen abundance and a mixing ratio of O/H$_{2} \geq (2.0\pm
0.6)\times 10^{-3}$ in Saturn's deep atmosphere. At lower
enrichments, there is not enough oxygen available to effectively
oxidize silicon and we would expect to see SiH$_{4}$ abundances
well above (up to five orders of magnitude) the observational
upper limit.

\section{Discussion}\label{Discussion}
\subsection{Constraints on Saturn's Oxygen Inventory}

The chemical constraints placed on the water and total oxygen
abundance on Saturn are summarized in Table 2. The independent
constraints given by CO and PH$_{3}$ indicate a water abundance of
$q\textrm{H}_{2}\textrm{O}\sim(1.7-5.5)\times10^{-3}$ in Saturn's
deep atmosphere, corresponding to a water enrichment of 1.9 to 6.1
times the solar system H$_{2}$O/H$_{2}$ ratio.  We therefore expect
water vapor to be the third or fourth most abundant gas (after
H$_{2}$, He, and CH$_{4}$) below the clouds of Saturn.

The silane mass balance constraint by itself shows that the total
oxygen abundance on Saturn must be enhanced by a factor of at
least 1.7 times protosolar. However, this constraint neglects that
after completely oxidizing Si and forming rock, enough oxygen must
be left over (as H$_{2}$O) to produce carbon monoxide in Saturn's
troposphere. But if too much oxygen remains, the observed amount
of PH$_{3}$ cannot form.  We therefore combine the CO \& SiH$_{4}$
constraints and the PH$_{3}$ \& SiH$_{4}$ constraints to estimate
the lower and upper limit, respectively, of Saturn's total oxygen
abundance (cf. Fegley \& Prinn 1988). Rewriting equation
(\ref{oxygenmassbalance}) using enrichment factors gives
\begin{equation}\label{oxygenmassbalancewithE}
E_{\textrm{\scriptsize{H}}_{2}\textrm{\scriptsize{O}}}(\textrm{O}_{\textrm{\scriptsize{H}}_{2}\textrm{\scriptsize{O}}})=E_{\textrm{\scriptsize{O}}}(\Sigma\textrm{O})-E_{\textrm{\scriptsize{rock}}}(\textrm{O}_{\textrm{\scriptsize{rock}}})
\end{equation}
where the terms in parentheses are the protosolar values. Replacing
these terms with values from Lodders (2003), equation
(\ref{oxygenmassbalancewithE}) can be rewritten as
\begin{equation}\label{enrichmentbalance}
E_{\textrm{\scriptsize{H}}_{2}\textrm{\scriptsize{O}}}=1.295E_{\textrm{\scriptsize{O}}}-0.295E_{\textrm{\scriptsize{rock}}}
\end{equation}
which relates the enrichment factors for water, oxygen, and rock,
and implicitly contains the silane mass balance constraint given in
equation (\ref{Simassconstraintsolar}) ($1.295/0.295\approx4.4$).
 Assuming $E_{\textrm{\scriptsize{rock}}}\approx7.4$ in equation
(\ref{enrichmentbalance}), the CO water constraint
($E_{\textrm{\scriptsize{H}}_{2}\textrm{\scriptsize{O}}}\geq1.9$)
gives an oxygen lower limit of $E_{\textrm{\scriptsize{O}}}\geq3.2$,
while the PH$_{3}$ water constraint
($E_{\textrm{\scriptsize{H}}_{2}\textrm{\scriptsize{O}}}\leq6.1$)
gives an oxygen upper limit of $E_{\textrm{\scriptsize{O}}}\leq6.4$.
 The upper and lower limits on the total oxygen abundance of Saturn are
summarized in Table 2.  Taken together, the chemical constraints
indicate a total oxygen enrichment in Saturn's interior of 3.2 to
6.4 times the solar system O/H$_{2}$ abundance, similar to that
observed for arsenic and less than the observed enrichments in
carbon and phosphorus.

From equation (\ref{oxygenmassbalancewithE}) it is evident that
estimates of the total oxygen enrichment depend on the assumed
rock enrichment on Saturn, so that smaller or larger rock
abundances require, respectively, smaller or larger enrichments in
oxygen. However, the observed CO and PH$_{3}$ mixing ratios along
with the general mass balance given in equation
(\ref{oxygenmassbalancewithE}) constrain the \emph{relative}
allowed abundances of water, oxygen, and rock over a range of rock
enrichments in Saturn's interior.

\subsection{Oxygen Partitioning into Metallic H-He}

The referee asked whether or not our chemical constraints on water
and oxygen are valid for Saturn as a whole or apply only to its
atmosphere. For example, one could ascribe the relative oxygen
depletion in Saturn's atmosphere to internal planetary
fractionation processes (e.g., see Fortney \& Hubbard 2003; 2004).
However, the selective or preferential removal of oxygen (with
respect to carbon and other heavy elements) from the molecular
envelope into a metallic core is difficult to model with current
knowledge about solution properties of C, N, O and other heavy
elements in metallic H-He. In fact the \emph{P}-\emph{T} curve for
the molecular to metallic hydrogen transition is unknown, as is
the high pressure phase diagram for the H-He system. The
hypothetical partitioning of oxygen into metallic H-He in Saturn's
interior is not our preferred model and we do not consider it
further.

Instead we think that our chemical constraints are valid for
Saturn as a whole. In support of this we note that the protosolar
H$_{2}$O/CH$_{4}$ ratio explains the observed atmospheric
chemistry of the brown dwarf Gliese 229b (e.g., Saumon et al.
2000). Although Gliese 229b apparently has subsolar metallicity
(Saumon et al. 2000) there is no evidence from chemical models
that the H$_{2}$O/CH$_{4}$ ratio itself is significantly smaller
than the protosolar value.

\subsection{The Nebular Snow Line}

Our chemical constraints on Saturn's water and total oxygen
inventory also have implications for Saturn's formation.  Our
results indicate that the
$E_{\textrm{\scriptsize{H}}_{2}\textrm{\scriptsize{O}}}/E_{\textrm{\scriptsize{CH}}_{4}}$
ratio ranges from 0.26 to 0.82 and the
$E_{\textrm{\scriptsize{O}}}/E_{\textrm{\scriptsize{C}}}$ ratio
ranges from 0.43 to 0.86 on Saturn (see Table 2). On Jupiter, the
observed methane abundance is 3.6 times protosolar while the
observed water abundance is 0.67 times protosolar, so that
$E_{\textrm{\scriptsize{H}}_{2}\textrm{\scriptsize{O}}}/E_{\textrm{\scriptsize{CH}}_{4}}\approx
0.19$ and
$E_{\textrm{\scriptsize{O}}}/E_{\textrm{\scriptsize{C}}}\approx
0.28$ (Lodders 2004).  In other words, both Jupiter and Saturn
apparently have lower H$_{2}$O/CH$_{4}$ and O/C ratios than
expected for uniform enrichments of carbon and oxygen (i.e.,
$E_{\textrm{\scriptsize{O}}}/E_{\textrm{\scriptsize{C}}}\equiv
1$). Lodders (2004) proposed that the water depletion and carbon
enrichment on Jupiter can be explained by moving the water ice
condensation front in the solar nebula (the nebular snow line)
beyond Jupiter's formation region. In this scenario, Jupiter never
accreted much water ice. Instead, a carbonaceous matter
condensation/evaporation front (the nebular tar line) was near
Jupiter's formation region and explains the carbon enrichment.
This scenario may also explain the subsolar H$_{2}$O/CH$_{4}$ and
O/C ratios that we have derived for Saturn, and account for the
large water enrichments derived for Uranus and Neptune by Lodders
\& Fegley (1994).  Further development of these ideas is beyond
the scope of this paper and will pursued elsewhere (Lodders \&
Fegley 2005, in preparation).

\section{Summary}\label{Summary}

We used CO, PH$_{3}$, and SiH$_{4}$ as chemical probes to provide
reliable estimates of the water and total oxygen abundances in the
deep atmosphere of Saturn. If the observed carbon monoxide mostly
forms in the troposphere, water must be enriched at least 1.9
times the solar system abundance.  The observed amount of
phosphine requires a water enrichment less than 6.1 times the
solar system abundance. The total oxygen abundance on Saturn must
be enhanced 3.2 to 6.4 times the protosolar O/H$_{2}$ ratio in
order to completely oxidize Si and form rock and yet leave an
appropriate amount of water to satisfy the CO and PH$_{3}$
constraints. Thus oxygen on Saturn appears to be enriched relative
to the solar system composition, but not to the same extent as
other heavy elements such as carbon and phosphorus.

\acknowledgments We thank K. Lodders for many helpful suggestions
and revisions and the anonymous referee for thoughtful comments.
This work was supported by NASA grant NAG5-11958.

\begin{figure}\label{figure1}
\plotone{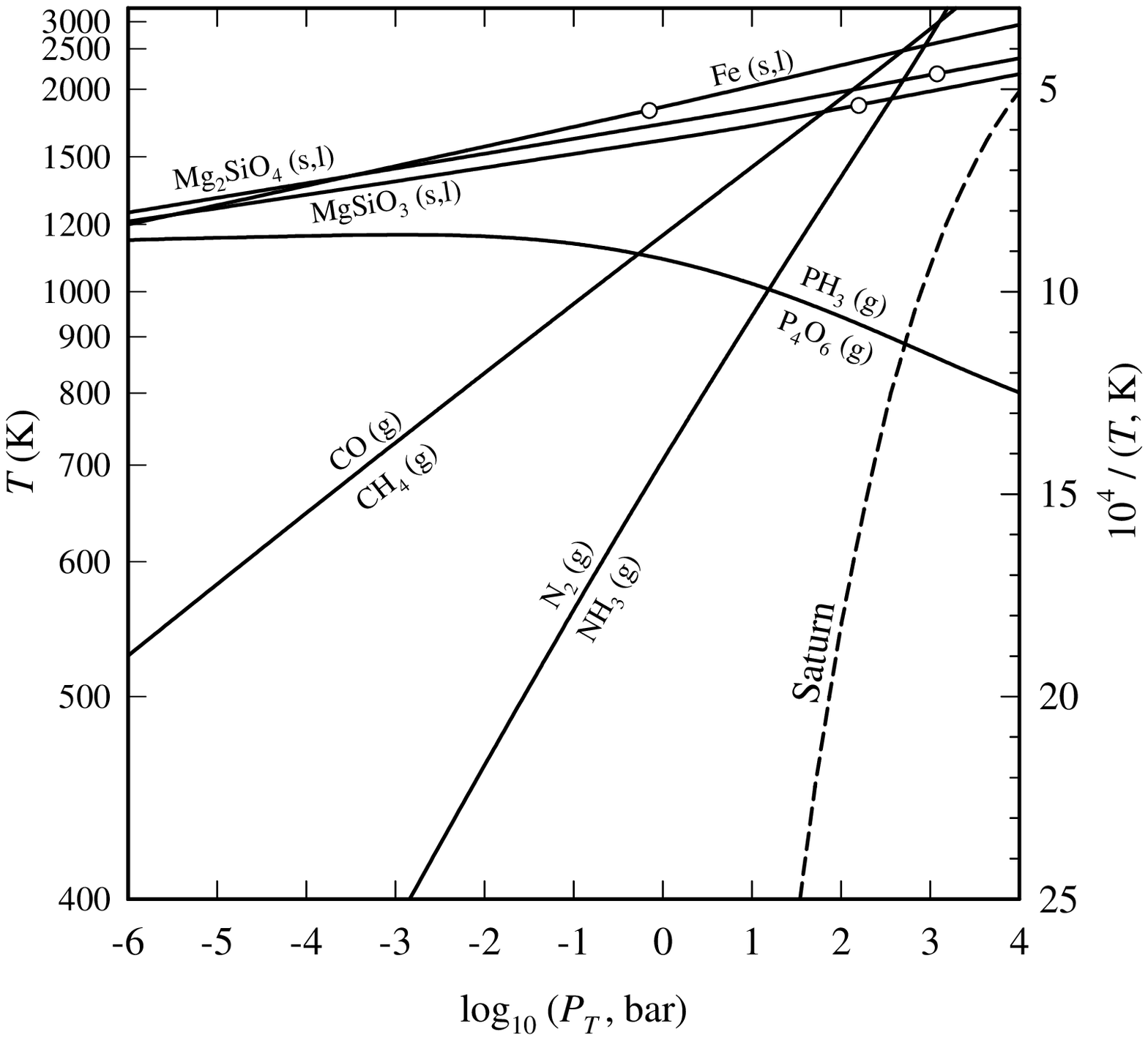} \caption{Equilibrium chemistry for a
protosolar-composition gas as a function of pressure and
temperature. The dashed line shows the location of the Saturnian
adiabat.  The lines labelled CO-CH$_{4}$, N$_{2}$-NH$_{3}$, and
PH$_{3}$-P$_{4}$O$_{6}$ indicate where the partial pressures of
the two gases are equal.  The lines labelled Fe,
Mg$_{2}$SiO$_{4}$, and MgSiO$_{3}$ show the condensation
temperatures of iron, forsterite, and enstatite as a function of
pressure, with circles denoting their melting temperatures.}
\end{figure}

\begin{figure}\label{figure2}
\plotone{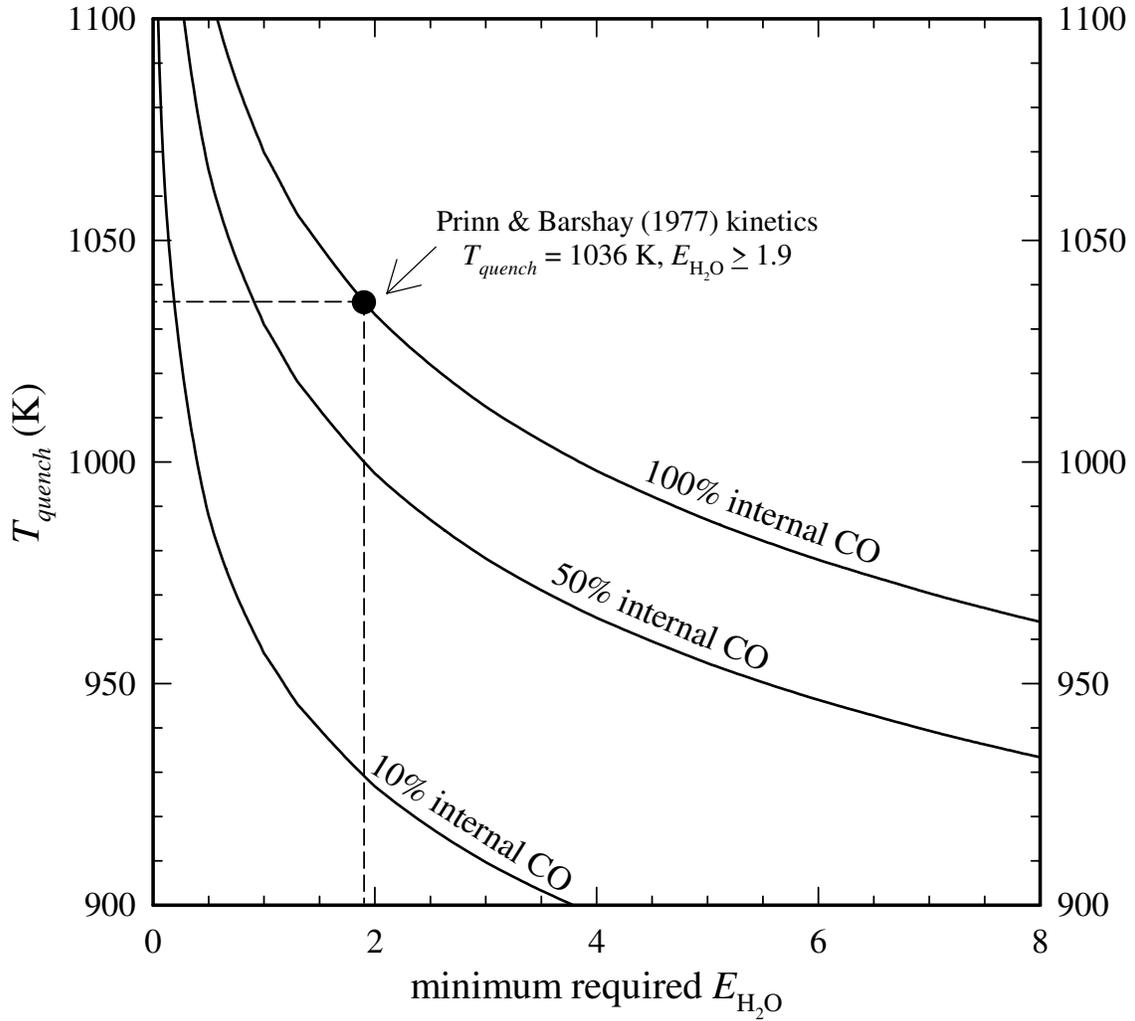} \caption{Relationship between the amount of CO
from an internal deep atmospheric source and the amount of water
required to produce the observed CO abundance. }
\end{figure}

\clearpage

\begin{deluxetable}{cccc}

\tablewidth{0pt}

\tablecaption{Gas Abundances in Saturn's Atmosphere} \label{Gas
Abundances Table}

\tablehead{\colhead{Gas} & \colhead{Observed Mixing
Ratio\tablenotemark{a}} & \colhead{Enrichment
Factor\tablenotemark{b}}} \startdata
H$_{2}$ & $\equiv$1 & $\equiv$1\\
He & 0.135 $\pm$ 0.025 & 0.700 $\pm$ 0.129\\
CH$_{4}$ & $(4.3\pm1.3)\times10^{-3}$ & 7.4 $\pm$ 2.3\\
SiH$_{4}$ & \textless(0.2-1.2)$\times10^{-9}$ & \textless10$^{-5}$-10$^{-6}$\\
GeH$_{4}$ & $(4.0\pm4.0)\times10^{-10}$ & 0.04 $\pm$ 0.04\\
NH$_{3}$ & $(1.8\pm1.3)\times10^{-4}$ & 1.1 $\pm$ 0.8\\
PH$_{3}$ & $(5.1\pm1.6)\times10^{-6}$ & 7.4 $\pm$ 2.3\\
AsH$_{3}$ & $(2.4\pm1.5)\times10^{-9}$ & 4.8 $\pm$ 3.0\\
H$_{2}$O & (2-200)$\times10^{-9}$ & 10$^{-4}$-10$^{-6}$\\
H$_{2}$S & \textless$0.4\times10^{-6}$ & \textless0.01\\
CO & $(1.6\pm0.8)\times10^{-9}$ & \nodata\\
\enddata
\tablenotetext{a}{Mixing ratios for a gas M (\emph{q}M) are
defined as M/H$_{2}$.} \tablenotetext{b}{Enrichment factor
($E_{\textrm{\scriptsize{M}}}$) for a gas M is defined as
(\emph{q}M)$_{\textrm{\scriptsize{Saturn}}}$/(\emph{q}M)$_{\textrm{\scriptsize{protosolar}}}$.}
\tablecomments{Saturn abundance data from Lodders \& \mbox{Fegley}
1998; Davis et al. 1996; de Graauw et al. 1997; Feuchtgruber et
al. 1997; Conrath \& Gautier 2000; Lellouch et al. 2001.  Solar
abundance data from Lodders 2003.}
\end{deluxetable}

\clearpage

\begin{deluxetable}{ccc}
\tablewidth{0pt} \tablecaption{Chemical Constraints on Saturn's
Water and Total Oxygen Abundance} \label{Chemical Constraints
Table} \tablehead{\colhead{Constraint} & \colhead{Mixing Ratio} &
\colhead{Enrichment Factor}}
\startdata \sidehead{Water}
CO & $\geq(1.7_{-0.4}^{+0.7})\times10^{-3}$ & $\geq1.9_{-0.5}^{+0.8}$\\
PH$_{3}$ & $\leq(5.5_{-2.5}^{+0.8})\times10^{-3}$ & $\leq6.1_{-2.7}^{+0.9}$\\
\sidehead{Total Oxygen} SiH$_{4}$ & $\geq(2.0\pm0.6)\times10^{-3}$
& $\geq1.7\pm0.5$\\ SiH$_{4}$ \& CO &
$\geq(3.7_{-0.5}^{+0.7})\times10^{-3}$ & $\geq3.2_{-0.4}^{+0.6}$\\
SiH$_{4}$ \& PH$_{3}$ & $\leq(7.4_{-2.4}^{+0.8})\times10^{-3}$ &
$\leq6.4_{-2.1}^{+0.7}$
\enddata
\tablecomments{Constraints on the water abundance are calculated
assuming $E_{\textrm{\scriptsize{CH}}_{4}} \approx
E_{\textrm{\scriptsize{PH}}_{3}} \approx 7.4\pm2.3$. Constraints
on the total oxygen abundance are calculated assuming a rock
enrichment of $E_{\textrm{\scriptsize{rock}}}\approx7.4\pm2.3$.}
\end{deluxetable}

\clearpage

\end{document}